\documentclass{mn2e}

\title[Distant future of Sun and Earth]{Distant future of the Sun and
       Earth revisited}
\author[K.-P. Schr\"oder and R.C. Smith]{K.-P. Schr\"oder$^{1}$\thanks{E-mail:
kps@astro.ugto.mx (KPS)} and Robert Connon Smith$^{2}$\thanks{E-mail:
R.C.Smith@sussex.ac.uk (RCS)}\\
$^{1}$Departamento de Astronom\'{\i}a, Universidad de Guanajuato, A.P.~144,
Guanajuato, C.P. 36000, GTO, M{\'e}xico\\
$^{2}$Astronomy Centre, Department of Physics and Astronomy, University of
Sussex, Falmer, Brighton BN1\,9QH, UK }
\begin{document}

\date{Accepted 2008 ....; Received 200~ ....;
      in original form 2007 September 25}

\pagerange{\pageref{firstpage}--\pageref{lastpage}} \pubyear{2008}

\maketitle

\label{firstpage}

\begin{abstract}
We revisit the distant future of the Sun and the solar system, based on
stellar models computed with a thoroughly tested evolution code.
For the solar giant stages, mass-loss by the cool (but not dust-driven)
wind is considered in detail. Using the new and well-calibrated mass-loss
formula of Schr\"oder \& Cuntz (2005, 2007), we find that the mass lost
by the Sun as an RGB giant (0.332 $M_{\odot}$, 7.59 Gy from now)
potentially gives planet Earth a significant orbital expansion,
inversely proportional to the remaining solar mass.

According to these solar evolution models, the closest encounter of planet
Earth with the solar cool giant photosphere will occur during the tip-RGB
phase. During this critical episode, for each time-step of the evolution
model, we consider the loss of orbital angular momentum suffered by planet
Earth from tidal interaction with the giant Sun, as well as dynamical drag in
the lower chromosphere. As a result of this, we find that planet Earth will
not be able to escape engulfment, despite the positive effect of solar
mass-loss. In order to survive the solar tip-RGB phase, any hypothetical planet
would require a present-day minimum orbital radius of about 1.15 AU.
The latter result may help to estimate the chances of finding planets
around White Dwarfs.

Furthermore, our solar evolution models with detailed mass-loss description
predict that the resulting tip-AGB giant will not reach its tip-RGB size.
Compared to other solar evolution models, the main reason is the more
significant amount of mass lost already in the RGB phase of the Sun.
Hence, the tip-AGB luminosity will come short of driving a final,
dust-driven superwind, and there will be no regular solar
planetary nebula (PN). The tip-AGB is marked by a last thermal pulse and
the final mass loss of the giant may produce a circumstellar (CS) shell
similar to, but rather smaller
than, that of the peculiar PN IC 2149 with an estimated total
CS shell mass of just a few hundredths of a solar mass.

\end{abstract}


\section{Introduction}\label{intro}
Climate change and global warming may have drastic effects on the
human race in the near future, over human time-scales of decades or
centuries. However, it is also of interest, and of relevance to the
far future of all living species, to consider the much
longer-term effects of the gradual heating of the Earth by a more
luminous Sun as it evolves towards its final stage as a white dwarf
star. This topic has been explored on several occasions
(e.g. Sackmann, Boothroyd \&\ Kraemer 1993, Rybicki \&\ Denis 2001,
Schr\"oder, Smith \&\ Apps 2001 (hereafter SSA)), and has been
discussed very recently by Laughlin (2007).

Theoretical models of solar evolution tell us that the Sun started on
the zero-age main sequence (ZAMS) with a luminosity only about 70\%\
of its current value, and it has been a long-standing puzzle that the
Earth seems none the less to have maintained a roughly constant
temperature over its life-time, in contrast to what an atmosphere-free
model of irradiation would predict. Part of the explanation may be
that the early atmosphere, rich in CO$_2$ that was subsequently locked
up in carbonates, kept the temperature up by a greenhouse effect which
decreased in effectiveness at just the right rate to compensate for
the increasing solar flux. The r\^ole of clouds, and their interaction
with galactic cosmic rays (CR), may also be important: there is now
some evidence (Svensmark 2007; but see Harrison et al. 2007 and Priest
et al.  2007) that cosmic rays encourage cloud cover at low altitudes,
so that a higher CR flux would lead to a higher albedo and lower
surface temperature. The stronger solar wind from the young Sun would
have excluded galactic cosmic rays, so cloud cover on the early Earth
may have been less than now, allowing the full effect of the solar
flux to be felt.

What of the future? Although the Earth's atmosphere may not be able to
respond adequately on a short time-scale to the increased greenhouse
effect of carbon dioxide and methane released into the atmosphere by
human activity, there is still the possibility, represented by James
Lovelock's Gaia hypothesis (Lovelock 1979, 1988, 2006), that the
biosphere may on a longer time-scale be able to adjust itself to
maintain life. Some doubt has been cast on that view by recent
calculations (Scaife, private communication, 2007; for details, see
e.g. Cox et al. 2004, Betts et al. 2004) which suggest that, on the
century timescale, the inclusion of biospheric processes in climate
models actually leads to an increase in carbon dioxide emissions,
partly through a feedback that starts to dominate as vegetation dies
back. In any case, it is clear that the time will come when the
increasing solar flux will raise the mean temperature of the Earth to
a level that not even biological or other feedback mechanisms can
prevent. There will certainly be a point at which life is no longer
sustainable, and we shall discuss this further in Section
\ref{habitablezone}.

After that, the fate of the Earth is of interest mainly insofar as it
tells us what we might expect to see in systems that we observe now at
a more advanced stage of evolution. We expect the Sun to end up as a
white dwarf -- do we expect there to be any planets around it, and in
particular do we expect any small rocky planets like the Earth?

The question of whether the Earth survives has proved somewhat tricky
to determine, with some authors arguing that the Earth survives
(e.g. SSA) and others (e.g. Sackmann et al. 1993) claiming that even
Venus survives, while general textbooks (e.g. Prialnik 2000, p.10)
tend to say that the Earth is engulfed. A simple model (e.g. SSA),
ignoring mass loss from the Sun, shows clearly that all the planets
out to and including Mars are engulfed, either at the red giant branch
(RGB) phase -- Mercury and Venus -- or at the later asymptotic giant
branch (AGB) phase -- the Earth and Mars. However, the Sun loses a
significant amount of mass during its giant branch evolution, and that
has the effect that the planetary orbits expand, and some of them keep
ahead of the advancing solar photosphere. The effect is enhanced by
the fact (SSA) that when mass loss is included the solar radius at the
tip of the AGB is comparable to that at the tip of the RGB,
instead of being much larger; Mars certainly survives, and it appears
(SSA) that the Earth does also.

The crucial question here is: what is the rate of mass loss in real
stars? Ultimately this must be determined from observations, but in
practice these must be represented by some empirical formula. Most people
use the classical Reimers' formula (Reimers 1975, 1977), but there is
considerable uncertainty in the value to be used for his parameter
$\eta$, and different values are needed to reproduce the observations
in different parameter regimes. In our own calculations (SSA) we used
a modification of the Reimers' formula, which has since been further improved
and calibrated rather carefully against observation, so that we believe
that it is currently the best available representation of mass loss
from stars with non-dusty winds (Schr\"oder \&\ Cuntz 2005, 2007 --
see Section \ref{evolution}, where we explore the consequences of this
improved mass-loss formulation).

However, although we have considerably reduced the uncertainties in the
mass-loss rate, there is another factor that works against the favourable
effects of mass loss: tidal interactions. Expansion of the Sun will cause
it to slow its rotation, and even simple conservation of angular momentum
predicts that by the time the radius has reached some 250 times its present
value (cf. Table \ref{solarmodels}) the rotation period of the Sun will have
increased to several thousand years instead of its present value of under a
month; effects of magnetic braking will lengthen this period even more.
This is so much longer than the orbital period of the Earth, even in its
expanded orbit, that the tidal bulge raised on the Sun's surface by the Earth
will pull the Earth back in its orbit, causing it to spiral inwards.

This effect was considered by Rybicki \&\ Denis (2001), who argued that Venus
was probably engulfed, but that the Earth might survive. An earlier paper by
Rasio et al. (1996) also considered tidal effects and concluded on the
contrary that the Earth would probably be engulfed. However, the Rybicki
\&\ Denis calculations were based on combining analytic representations of
evolution models (of Hurley, Pols \&\ Tout 2000) with the original
Reimers' mass-loss formula rather than on full solar evolution calculations
with a well-calibrated mass-loss formulation. The Rasio et al. paper also
employed the original Reimers' formula, and both papers use somewhat
different treatments of tidal drag.  We have therefore re-considered this
problem in detail, with our own evolutionary calculations and an improved
mass-loss description as the basis; full details are given in Sections
\ref{evolution} and \ref{tiprgb}.

\section{Solar evolution model with mass loss}\label{evolution}

In order to describe the long-term solar evolution, we use the Eggleton
evolution code (Eggleton 1971, 1972, 1973) in the version described by
Pols et al. (1995, 1998), which has updated opacities and an improved
equation of state. Among other desirable characteristics, his code uses
a self-adapting mesh and a $\nabla$-based prescription of ``overshooting'',
which has been well-tested and calibrated with giant stars in
eclipsing binaries (for details, see Schr\"oder et al. 1997, Pols et al.
1997, Schr\"oder 1998). Because of the low mass and a non-convective
core, solar evolution models are, however, not subject to any MS
(main sequence) core-overshooting. In use, the code is very fast, and
mass-loss is accepted simply as an outer boundary condition.

As already pointed out by VandenBerg (1991), evolution codes have the
tendency to produce, with their most evolved models, effective temperatures
that are slightly higher than the empirically determined values.
The reason lies, probably, in an inadequacy of both low-temperature
opacities and mixing-length theory at low gravity. With the latter, we
should expect a reduced efficiency of the convective energy transport for
very low gravity, because the largest eddies are cut out once the ratio of
eddy-size to stellar radius has increased too much with $g^{-1}$.
Hence, as described by Schr\"oder, Winters \& Sedlmayr (1999), our
mixing-length parameter, normally $\alpha = 2.0$ for $\log{g}<1.94$,
receives a small adjustment in the form of a gradual reduction for
supergiant models, reaching $\alpha = 1.67$ at $\log{g} = 0.0$.
With this economical adjustment, our evolution models now give
a better match to empirically determined effective
temperatures of very evolved late-type giants and supergiants, such
as $\alpha^1$ Her (see Schr\"oder \& Cuntz 2007, Fig. 4 in particular),
and even later stages of stellar evolution (Dyck et al. 1996,
and van Belle et al. 1996, 1997).

The evolution model of the Sun presented here uses an opacity grid that
matches the empirical solar metallicity of Anders \& Grevesse (1989),
$Z$ = 0.0188, derived from atmospheric models with simple 1D radiative
transfer -- an approach consistent with our evolution models.
Together with $X$ = 0.700 and $Y$ = 0.2812, there is a good match 
with present-day
solar properties derived in the same way (see Pols et al. 1995). We note
that the use of 3D-hydrodynamic modelling of stellar atmospheres and their
radiative transfer may lead to a significantly lower solar abundance scale
(e.g., Asplund, Grevesse \& Sauval  2005, who quote $Z$ = 0.0122), but these
lower values are still being debated, and create some problems with
helioseismology.
Of course, using lower metallicities with an evolution code always results
in more compact and hotter stellar models. Hence, if we used a lower $Z$
our code would plainly
fail to reproduce the present-day Sun, and the reliability of more
evolved models with lower $Z$ must therefore also be seriously doubted.

The resulting solar evolution model suggests an age of the present-day MS
Sun of 4.58\,Gy ($\pm 0.05$\,Gy), counted from its zero-age MS start model,
which is well within the range of commonly accepted values for the real age
of the Sun and the solar system (e.g. Sackmann et al. 1993). Our model also
confirms some well-established facts: (1) The MS-Sun has already undergone
significant changes, i.e., the present solar luminosity $L$ exceeds
the zero-age value by 0.30\,$L_{\odot}$, and the zero-age solar radius
$R$ was 11\% smaller than the present value. (2) There was an increase
of effective temperature $T_{\rm eff}$ from, according to our model,
5596\,K to 5774\,K ($\pm 5$\,K). (3) The present Sun is increasing its
average luminosity at a rate of 1\% in every 110 million years, or
10\% over the next billion years. All this is completely consistent with
established solar models like the one of Gough (1981).

Certainly, the solar MS-changes and their consequences for Earth are extremely
slow, compared to the current climate change driven by human factors.
Nevertheless, solar evolution will force global warming
upon Earth already in the ``near'' MS future of the Sun, long before
the Sun starts its evolution as a giant star (see our discussion of the
habitable zone in Section \ref{habitablezone}).

At an age of 7.13\,Gy, the Sun will have reached its highest
$T_{\rm eff}$ of 5820\,K, at a luminosity of 1.26\,$L_{\odot}$.
From then on, the evolving MS Sun will gradually become cooler,
but its luminosity will continue to increase. At an age of 10.0\,Gy,
the solar effective temperature will be back at $T_{\rm eff} = 5751$\,K,
while $L = 1.84\,L_{\odot}$, and the solar radius then will be 37\% larger
than today. Around that age, the evolution of the Sun will speed up,
since the solar core will change from central hydrogen-burning
to hydrogen shell-burning and start to contract. In response, the
outer layers will expand, and the Sun will start climbing up the RGB
(the ``red'' or ``first giant branch'' in the HRD) -- at first very gradually,
but then accelerating. At an age of 12.167\,Gy, the Sun will have
reached the tip of the RGB, with a maximum luminosity of 2730\,$L_{\odot}$.

\begin{table}
 \begin{minipage}{140mm}
  \caption{Main physical properties of characteristic solar
models}\label{solarmodels}
  \begin{tabular}{lccccc}
\hline
Phase & Age/Gy & $L/L_{\odot}$ & $T_{\rm eff}/$K & $R/R_{\odot}$ &
$M_{\rm Sun}/M_{\odot}$\\
\hline
ZAMS       &  0.00 & 0.70 & 5596 & 0.89 & 1.000 \\
present    &  4.58 & 1.00 & 5774 & 1.00 & 1.000 \\
MS:hottest &  7.13 & 1.26 & 5820 & 1.11 & 1.000 \\
MS:final   & 10.00 & 1.84 & 5751 & 1.37 & 1.000 \\
RGB:tip    & 12.17 & 2730.& 2602 & 256. & 0.668 \\
ZA-He      & 12.17 & 53.7 & 4667 & 11.2 & 0.668 \\
AGB:tip    & 12.30 & 2090.& 3200 & 149. & 0.546 \\
AGB:tip-TP & 12.30 & 4170.& 3467 & 179. & 0.544 \\
\hline
\end{tabular}
\end{minipage}
{\footnotesize (note: 1.00 AU = 215 $R_{\odot}$)}
\end{table}

In order to quantify the mass-loss rate of the evolved, cool solar
giant at each time-step, we use the new and well-calibrated mass-loss formula
for ordinary cool winds (i.e., not driven by dust) of Schr\"oder \& Cuntz
(2005, 2007). This relation is, essentially, an improved Reimers' law,
physically motivated by a consideration of global chromospheric properties
and wind energy requirements:
\begin{equation}
\dot{M} \ = \ \eta  \frac{L_* R_*}{M_*}
{\Bigl(\frac{T_{\rm eff}}{4000\,{\rm K}}\Bigr)}^{3.5}
\Bigl(1+\frac{g_{\odot}}{4300\,g_*}\Bigr)
\end{equation}
with $\eta = 8 \times 10^{-14} M_\odot$ y$^{-1}$, \ $g_\odot$
= solar surface gravitational acceleration, and
$L_*$, $R_*$, and $M_*$ in solar units.

This relation was initially calibrated by Schr\"oder \& Cuntz (2005)
with the total mass loss on the RGB, using the blue-end (i.e., the least
massive) horizontal-branch (HB) stars of globular clusters with different
metallicities. This method avoids the interfering problem of temporal
mass-loss variations found with individual giant stars and leaves an
uncertainty of the new $\eta$-value of only 15\%, just under the individual
spread of RGB mass-loss required to explain the width of HBs.

Later, Schr\"oder \& Cuntz (2007) tested their improved mass-loss relation
with six nearby galactic giants and supergiants, in comparison with
four other, frequently quoted mass-loss relations. All but one
of the tested giants are AGB stars, which have (very different)
well-established physical properties and empirical mass-loss rates, all
by cool winds {\it not} driven by radiation-pressure on dust.
Despite the afore-mentioned problem with the inherent time-variability
of this individual-star-approach, the new relation (equation~(1))
was confirmed to give the best representation of the cool, but not
``dust-driven'' stellar mass-loss: it was the only one that
agreed within the uncertainties (i.e., within a factor of 1.5 to 2) with
the empirical mass-loss rates of {\it all} giants. Hence, since
the future Sun will not reach the critical luminosity required
by a ``dust-driven'' wind (see Section \ref{tipagb}), we here
apply equation~(1) to describe its AGB mass-loss as well as its RGB mass-loss.

\begin{figure}
\vspace{60mm}
\includegraphics{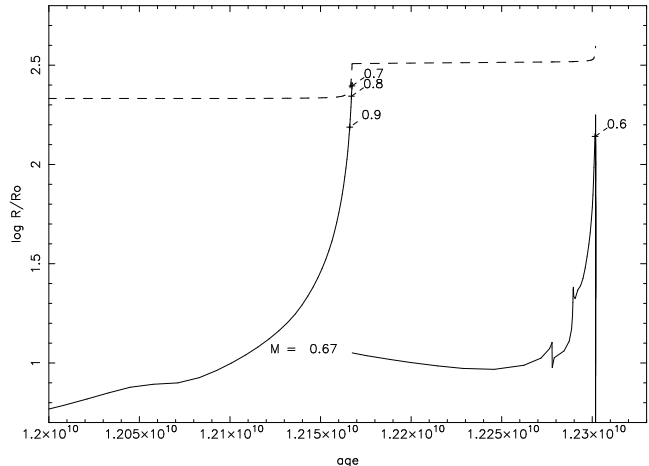}
\caption{Solar radius evolution during the RGB and AGB phases. Included for
         comparison (dashed curve) is the potential orbital radius of
         planet Earth, taking account of solar mass loss but neglecting
         any loss of orbital angular momentum. The labels on the curve for
         the solar radius show the mass of the Sun in units of its
         present-day mass.}\label{radius}
\end{figure}

The exact mass-loss suffered by the future giant Sun has, of
course, a general impact on the radius of the solar giant, since
the reduced gravity allows for an even larger (and cooler) supergiant.
The luminosity, however, is hardly affected because it is mostly set
by the conditions in the contracting core and the hydrogen-burning shell.
In total, our solar evolution model yields a loss of $0.332\,M_{\odot}$ by
the time the tip-RGB is reached (for $\eta =
8 \times 10^{-14} M_{\odot} y^{-1}$). This is a little more than the
$0.275\,M_{\odot}$ obtained
by Sackmann et al. (1993), who used a mass-loss prescription based on the
original, simple Reimers' relation. Furthermore, our evolution model
predicts that at the very tip of the RGB, the Sun should reach
$R = 256\,R_{\odot} = 1.2$\,AU (see Fig. \ref{radius}),
with $L = 2730 L_{\odot}$ and $T_{\rm eff} = 2602$\,K. More details are given 
in Table \ref{solarmodels}.

By comparison, a prescription of the (average) RGB mass-loss rate with
$\eta = 7 \times 10^{-14} M_{\odot} $y$^{-1}$, near the lower error limit
of the mass-loss calibration with HB stars, yields a solar model
at the very tip of the RGB with $R = 249 R_{\odot}$, $L = 2742 L_{\odot}$,
$T_{\rm eff} = 2650$\,K, and a total mass lost on the RGB of
$0.268 M_{\odot}$. With $\eta = 9 \times 10^{-14} M_{\odot} $y$^{-1}$,
on the other hand, the Sun would reach the very tip of the RGB with
$R = 256 R_{\odot}$, $L = 2714 L_{\odot}$, $T_{\rm eff} = 2605$\,K,
and will have lost a total of $0.388 M_{\odot}$.
While these slightly different possible outcomes of solar tip-RGB evolution
 -- within the uncertainty of the mass-loss prescription -- require
further discussion, which we give in Section \ref{doomsday}, the differences are
too small to be obvious on the scale of Fig. \ref{radius}.

With the reduced solar mass and, consequently, lower gravitational attraction,
all planetary orbits -- that of the Earth included -- are bound to expand.
This is simply a consequence of the conservation of angular momentum
$\Lambda_E = M_E \ v_E \ r_E$, while the orbital radius
(i.e. $r_E$) adjusts to a new balance between
centrifugal force and the reduced gravitational force of the Sun,
caused by the reduced solar mass $M_{\rm Sun}(t)$. Substituting $v_E =
\sqrt{G M_{\rm Sun}(t)/r_E}$
in $\Lambda_E$ yields $r_E \propto \Lambda_E^2/M_{\rm Sun}(t)$. For this 
conservative case, we find that $r_E$ is 1.50\,AU for the case $\eta =
8 \times 10^{-14} M_{\odot} y^{-1}$. For the smaller ($7 \times 10^{-14}$) and
larger ($9 \times 10^{-14}$) values of $\eta$, we find, respectively, 
$r_E = 1.37$\,AU and $r_E = 1.63$\,AU, so in all cases the orbital radius 
is comfortably more than
the solar radius, when angular momentum is conserved.

Section \ref{tides} provides a treatment of the more realistic case,
in which angular momentum is {\it not} conserved. We have taken great care
in determining the mass-loss and other parameters for our models, because
the best possible models of the evolution of solar mass and radius through
the tip-RGB phase are required to provide reliable results.

The significant solar RGB mass loss will also shape the later solar
AGB evolution. Compared with models without mass loss, the AGB Sun will not
become as large and luminous, and will be shorter-lived, because it
lacks envelope mass for the core and its burning shells to ``eat'' into.
In fact, the solar tip-AGB radius ($149\,R_{\odot}$) will never reach
that of the tip-RGB (see Fig. \ref{radius}), and AGB thermal pulses
are no threat to any planet which would have survived the tip-RGB. Our
evolution code resolved only the two final and most dramatic thermal
pulses (cf. Section \ref{tipagb}).

The regular tip-AGB luminosity of $2090\,L_{\odot}$ will not exceed
the tip-RGB value, either. Hence, as will be discussed in
Section \ref{tipagb}, the tip-AGB Sun will not develop a sustained
dust-driven superwind but will stay short of the critical luminosity
required by dust-driven winds (see Schr\"oder et al. 1999). The very
tip of the AGB coincides with a thermal pulse (TP), after which the
giant briefly reaches a peak luminosity of 4170\,$L_{\odot}$, but at a
higher $T_{\rm eff} = 3467$\,K than on the RGB (see Table \ref{solarmodels}
and Section \ref{tipagb}), keeping
the radius down to $179\,R_{\odot}$. Again, the best possible treatment of all
prior mass loss from the giant Sun is essential for modelling this phase
reliably.

\section{Evolution of the habitable zone}\label{habitablezone}

The Earth currently sits in the `habitable zone' in the solar system,
that is, the region in which conditions on the Earth -- in particular
the average planetary temperature -- are favourable for life.
There are various precise definitions of `habitability' in the literature,
and a useful overview of habitable zones in the wider context of
extrasolar planetary systems is given by Franck et al. (2002).
For the current paper, a convenient definition is that a planet is
habitable if the conditions on it allow the presence of liquid water
on its surface. This may allow extremes of temperature that would make
life uncomfortable if not impossible for humans, but the argument is that
life of any kind (at least any kind we know about at present) requires water
at some stage in its life cycle. We shall adopt that definition in this paper,
but note that even with that apparently simple definition it is not
straightforward to calculate the width of the habitable zone.

It may be instructive to begin with a calculation of the mean planetary
temperature in terms of a spherical black body, by assuming that the
planetary body absorbs the solar flux intercepted by its (circular)
cross-sectional area and re-emits it spherically symmetrically at a black
body temperature $T$. Then (cf. SSA) $T$ is given by
\begin{eqnarray}
T  & = & (1-A)^{1/4} \left( \frac{R}{2D} \right)^{1/2} T_{\rm eff} \nonumber \\
   & = &
0.0682\,(1-A)^{1/4} \left( \frac{R}{R_{\odot}}\right)^{1/2}
\left(\frac{1 \rm{AU}}{2D}\right)^{1/2} T_{\rm eff} \label{Tbb}
\end{eqnarray}
where $D$ is the distance of the body from the centre of the Sun, $R$ is the
radius of the Sun, $A$ is the Bond albedo of the Earth and $T_{\rm eff}$ is the
effective temperature of the Sun. On that basis, taking $T_{\rm eff}$ = 5774\,K
and $R = R_{\odot}$ (Table \ref{solarmodels}), and $A = 0.3$ (Kandel \&\
Viollier 2005), we find $T(1\,{\rm AU})$ = 255\,K. But the actual mean
temperature of the Earth at present is 33\,K warmer, at $T=288$\,K.
This demonstrates the warming effect of our atmosphere, which becomes
significantly more important with higher temperature (see below).

In fact, there are various complex, partly antagonistic atmospheric
feedback mechanisms (for example, the greenhouse effect, the variation
of planetary albedo with the presence of clouds, snow and ice, and the
carbonate-silicate cycle which determines the amount of carbon dioxide
in the atmosphere) that act to change the surface temperature from what
it would be in the absence of an atmosphere. These mechanisms have been
carefully discussed by Kasting, Whitmire \& Reynolds (1993), who conclude
that a conservative estimate of the current habitable zone (HZ) stretches
from 0.95\,AU to 1.37\,AU. We shall adopt their result for the limited
purposes of this paper. It can be adjusted in a simple-minded way to allow
for the evolution of the Sun by scaling the inner and outer HZ radii
$r_{\rm HZ,i}$, $r_{\rm HZ,o}$ with the changing solar luminosity
$L_{\rm Sun}(t)$: $r_{\rm HZ} \propto \sqrt{L_{\rm Sun}(t)}$. In this way,
the respective critical values of solar irradiance derived by Kasting et
al.~(1993) for the inner and outer edge of the HZ are maintained.

Certainly, with the 10\% increase of solar luminosity over the next 1\,Gy
(see previous section), it is clear that Earth will come to leave the HZ
already in about a billion years time, since the inner (hot side) boundary
will then cross 1\,AU. By the time the Sun comes to leave the main sequence,
around an age of 10\,Gy (Table \ref{solarmodels}), our simple model predicts
that the HZ will have moved out to the range 1.29 to 1.86\,AU. The Sun will
have lost very little mass by that time, so the Earth's orbital radius will
still be about 1\,AU -- left far behind by the HZ, which will instead be
enveloping the orbit of Mars.

By the time the Sun reaches the tip of the RGB, at 12.17\,Gy, the Earth's
orbital radius will only have expanded to at most 1.5\,AU, but the habitable
zone will have a range of 49.4 to 71.4\,AU, reaching well into the Kuiper
Belt! The positions of the HZ boundaries are not as well determined as
these numbers suggest, because in reality the scaling for the boundaries
of the HZ almost certainly depends also on how clouds are affected by
changes in the solar irradiance. These effects are complex and uncertain
(cf. Kasting 1988), and may increase or decrease the speed at which the
HZ drifts outwards. But none the less it seems clear that the HZ will move
out past the Earth long before the Sun has expanded very much, even if
the figure of one billion years is a rather rough estimate of how long
we have before the Earth is uninhabitable.

In other planetary systems around solar-type stars, conditions may be
different, and it may even be possible for life to start during a star's
post-main-sequence evolution, if a planet exists at a suitable distance from
the star. This possibility is discussed by Lopez, Schneider \&\ Danchi (2005),
who also give a general discussion of the evolution of habitable zones with
time. However, they use the evolution models of Maeder \&\ Meynet (1988),
which do not agree as well as ours with the colours and observed $T_{\rm eff}$'s
of the red giants in star clusters (see, e.g., illustrations given by
Meynet et al. 1993), and which predict a very
different behaviour for the solar radius; so their results are not
directly comparable with ours.

What will happen on the Earth itself? Ignoring for the moment the
short-timescale (decades to centuries) problems currently being introduced
by climate change, we may expect to have about one billion years of time
before the solar flux has increased by the critical 10\% mentioned earlier.
At that point, neglecting the effects of solar irradiance changes on the
cloud cover, the water vapour content of the atmosphere will increase
substantially and the oceans will start to evaporate (Kasting 1988). An
initially moist greenhouse effect (Laughlin 2007) will cause runaway
evaporation until the oceans have boiled dry. With so much water vapour in
the atmosphere, some of it will make its way into the stratosphere. There,
solar UV will dissociate the water molecules into OH and free atomic hydrogen,
which will gradually escape, until most of the atmospheric water vapour has
been lost. The subsequent dry greenhouse phase will raise the surface
temperature significantly faster than would be expected from our very simple
black-body assumption, and the ultimate fate of the Earth, if it survived
at all as a separate body (cf. Section \ref{tiprgb}), would be to become
a molten remnant.

\section{The inner planetary system during tip-RGB evolution}\label{tiprgb}
After 12 Gy of slow solar evolution, the final ascent of the RGB will
be relatively fast. The solar radius will sweep through the inner planetary
system within only 5 million years, by which time the evolved solar giant will
have reached the tip of the RGB and then entered its brief (130 million year)
He-burning phase. The giant will first come to exceed the orbital size
of Mercury, then Venus. By the time it approaches Earth, the solar mass-loss
rate will reach up to $2.5 \times 10^{-7} M_{\odot}$\,y$^{-1}$ and lead to
some orbital expansion (see Section \ref{evolution}).
But the extreme proximity of
the orbiting planet to the solar photosphere requires the consideration
of two effects, which both lead to angular momentum loss and a fatal
decrease of the orbital radius of planet Earth.

\subsection{Tidal interaction}\label{tides}
For the highly evolved giant Sun, we may safely assume (cf. Section
\ref{intro}) that it has essentially ceased to rotate, after nearly 2 billion
years of post-MS magnetic braking acting
on the hugely expanded, cool RGB giant. Consequently, any tidal interaction
with an orbiting object will result in its suffering a continuous drag by the
slightly retarded tidal bulges of the giant solar photosphere.

As shown in Section \ref{evolution}, the orbital radius of planet Earth
$r_E$ depends on the angular momentum squared, by the equation
\begin{equation}
r_E = \frac{\Lambda^2_E(t)}{M_E^2 \, G \, M_{\rm Sun}(t)} .
\label{r_e}
\end{equation}
Hence, the terrestrial orbit reacts quite sensitively to any
loss of angular momentum, by shrinking.

The retardation of the tidal bulges of the solar photosphere will
be caused by tidal friction in the outer convective envelope of the
RGB Sun. This physical process was analyzed, solved and applied
by J.-P. Zahn (1977, 1989, and other work referred to therein),
and successfully tested with the synchronization and circularization
of binary star orbits by Verbunt \& Phinney (1995). In a convective envelope,
the main contribution to tidal
friction comes from the retardation of the equilibrium tide by
interaction with convective motions. For a circular orbit, the
resulting torque $\Gamma$ exerted on planet Earth by the retarded solar
tidal bulges is given by (Zahn 1977;  Zahn 1989, Eq.11):
\begin{equation}
\Gamma = 6 \frac{\lambda_2}{t_f} q^2 M_{\rm Sun} R_{\rm Sun}^2 \Bigl(
\frac{R_{\rm Sun}}{r_E} \Bigr)^6 (\Omega - \omega) .
\label{gamma-tidal}
\end{equation}
Here, the angular velocity of the solar rotation is supposed to be $\Omega
 =0$, while that of the orbiting Earth, $\omega(t) = 2\pi/P_E(t) =
\Lambda^{-3}(t)
M^3_E (G M_{\rm Sun}(t))^2$, will vary both with the decreasing angular
momentum $\Lambda(t)$ ($= 2.67\times 10^{40}$kg\,m$^2$\,s$^{-1}$ at present)
and with
the solar mass in the final solar RGB stages. The exerted torque scales with
the square of the (slowly increasing) mass ratio $q(t) = M_E/M_{\rm Sun}(t)$
($= 3.005 \times 10^{-6}$ at present), because $q$ determines the magnitude
of the tidal bulges.
$t_f(t) = (M_{\rm Sun}(t) R_{\rm Sun}^2(t)/L_{\rm Sun}(t))^{1/3}
\approx $ O(1\,y) is the convective friction time (Zahn 1989, Eq.7),
and the coefficient $\lambda_2$ depends on the properties of the convective
envelope. For a fully convective envelope (Zahn 1989, Eq.15), with a tidal
period $\approx$ O(1\,y), comparable to $2t_f$, we may use
$\lambda_2 \approx 0.019 \,\alpha^{4/3}
\approx 0.038$ (with a convection parameter of our tip-RGB solar model
of $\alpha \approx 1.7$). This coefficient appears to be the main
source of uncertainty (see Section \ref{doomsday}), because it is related
to the simplifications of the mixing length theory (MLT).

With the properties of the tip-RGB Sun, a typical value of the tidal drag
acting on planet Earth is $\Gamma = d\Lambda/dt =
 - 3.3 \times 10^{26}$kg\,m$^2$\,s$^{-2}$, which gives a typical
orbital angular momentum decay time of
$\tau = \mid\Lambda / \Gamma\mid = 2.6 \times 10^6$\,y. This is comparable to
the time spent by the Sun near the tip-RGB;
since a loss of only $\approx 10$\% of the angular momentum
will be sufficient to reduce the orbital radius (by 20\%) to lower it
into the solar giant photosphere, this order-of-magnitude calculation
illustrates clearly that tidal interaction is crucial. Its
full consideration requires a timestep-by-timestep computation of
the loss of orbital angular momentum; at each time-step of the solar evolution
calculation, we use equation~(\ref{gamma-tidal}), together with the radii and
masses of our solar evolution model, to compute the change in angular momentum,
and then use equation (\ref{r_e}) to compute the change in the orbital radius, and
hence the new orbital period of the Earth.
Section \ref{doomsday} presents the result, which also takes into account
the relatively small additional angular momentum losses by dynamical drag, as
discussed in the next section.

\subsection{Dynamical friction in the lower chromosphere}\label{drag}
A further source of angular momentum loss by drag is dynamical friction,
from which any object suffers in a fairly close orbit, by its
supersonic motion through the gas of the then very extended, cool solar
giant chromosphere.
In a different context, dynamical drag exerted by a giant atmosphere
has already been considered by Livio \& Soker (1984).
But the specific problem here is to find
an adequate description of the density structure of the future cool solar
giant. Fortunately, as it turns out (see below),
dynamical drag will play only a minor r{\^o}le, very near the
solar giant photosphere, and the total angular momentum loss
is dominated by the tidal interaction described above. An approximate
treatment of the drag is therefore adequate, and we use the recent study
by Ostriker (1999).

In the case of supersonic motion (with a Mach number\footnote{Note that
$v_{\rm E} \propto M_{\rm Sun}(t)$, and so the Mach number is somewhat
lower than would be expected from the present orbital velocity of the
Earth of about 30\,km\,s$^{-1}$.} of the order of 2 to 3)
in a gaseous medium, dynamical friction consists in about equal shares
of the collisionless, gravitational interaction with its wake and of the
friction itself. In her study, Ostriker (1999, Fig. 3) finds that
the drag force exerted on the object in motion is

\begin{equation}
F_d = \lambda_d \, 4\pi  \rho \, (G M_E/c_s)^2
\label{f-drag}
\end{equation}
where $\lambda_d$ is of the order of 1 to 3. The numerical simulations made by
S{\'a}nchez-Salcedo \& Brandenburg (2001) are in general agreement with the
results of Ostriker (1999). Here $c_s$ is the speed of sound, which in a 
stellar chromosphere is about 8\,km\,s$^{-1}$, and $\rho$ is the gas density 
(SI units). The latter quantity
is the largest source of uncertainty, as we can only make guesses (see below)
as to what the gas density in the lower giant solar chromosphere will be.
The angular momentum loss resulting from this drag is simply
\begin{equation}
d\Lambda/dt = - F_d \, r_E,
\label{gamma-drag}
\end{equation}
and the corresponding life-time of the orbital
angular momentum is $\tau = \Lambda/\mid d\Lambda/dt\mid$, as above.

For the lower chromosphere of the K supergiant $\zeta$~Aur,
employing an analysis of the additional line absorption in the spectrum of
a hot companion in chromospheric eclipse, Schr\"oder, Griffin \&\ Griffin
(1990) found an average hydrogen particle density of
$7 \times 10^{11}$\,cm$^{-3}$ at a height of $2 \times 10^6$\,km.
Alternatively, we may simply assume that the density of the lower solar
chromosphere scales with gravity $g$, which will be lower by 4.7
orders of magnitude on the tip-RGB, while the density scale-height scales with
$g^{-1}$ (as observations of cool giant chromospheres seem to indicate,
see Schr\"oder 1990). The chromospheric models of both
Lemaire et al. (1981) and Maltby et al. (1986) suggest particle
densities of the order of $10^{17}$\,cm$^{-3}$ at a height of 100\,km,
and a scale height of that order for the present, low solar
chromosphere. Scaled to tip-RGB gravity, that would correspond to a
particle density of $2 \times 10^{12}$\,cm$^{-3}$,
or $\rho \approx 4 \times 10^{-9}$\,kg\,m$^{-3}$,
at a height of $5 \times 10^6$\,km (0.03 AU), and a density scale height
of that same value.

For the computation of the orbital angular momentum loss of the Earth, presented
below (see Figures \ref{drag_earth} and \ref{drag_crit}), we apply the
latter, rather higher values of the future
chromospheric gas density, together with the (also more pessimistic)
assumption of $\lambda_d = 3$ (using $c_s = 8$\,km\,s$^{-1}$).
The typical angular momentum decay-time by dynamical friction in the
low ($h \approx 0.03$\,AU) chromosphere of the tip-RGB solar giant is
14 million years  -- significantly longer than that for tidal
interaction. Hence, this illustrates that dynamical friction is of
interest only in the lowest chromospheric
layers, adding there just a little to the drag exerted by
tidal interaction. None the less, we include it, using equations (\ref{f-drag})
and (\ref{gamma-drag}) to calculate the additional angular momentum change to be
included in equation (\ref{r_e}).

\subsection{``Doomsday'' confirmed}\label{doomsday}
\begin{figure}
\vspace{60mm}
\includegraphics{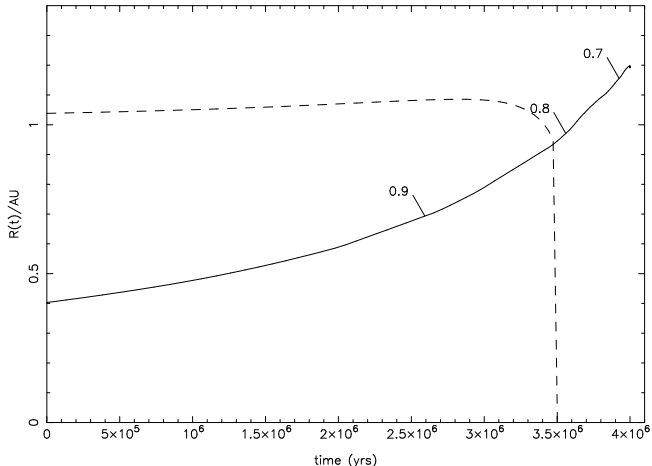}
\caption{The final 4 million years of solar evolution before the tip-RGB,
showing the radii of the Sun and of the orbit of planet Earth (dashed curve)
 -- taking account of angular momentum losses by tidal
interaction and supersonic drag. The labels on the solar radius track
give values of $M_{\rm Sun}(t)/M_\odot$, as in Figure
\ref{radius}.}\label{drag_earth}
\end{figure}

As explained in the previous two sections, we use equations (\ref{r_e})
to (\ref{gamma-drag}) to compute, at each time-step of our evolutionary calculation,
a detailed description of the orbital evolution for planet
Earth in the critical tip-RGB phase of the Sun under the
influence of tidal interaction and dynamical drag. The resulting evolution both of the
orbital radius of the Earth and of the radius of the solar giant is shown in
Fig. \ref{drag_earth}. This shows that, despite the reduced gravity from a less
massive tip-RGB Sun,
the orbit of the Earth will hardly ever come to exceed 1\,AU by a significant
amount. The potential orbital growth given by the reduced solar mass
is mostly balanced and, eventually, overcome by the effects of tidal
interaction. Near the very end, supersonic drag also becomes a significant
source of angular momentum loss.

As shown by Fig. \ref{drag_earth}, engulfment and loss of planet Earth
will take place just before the Sun reaches the tip of the RGB, 7.59\,Gy
($\pm 0.05$\,Gy) from now. According to our calculation, it occurs when
the RGB Sun has still another 0.25\,AU to grow, about
500,000 years before the tip-RGB. Of course, Mercury and Venus will
already have suffered the same fate as Earth some time before --
respectively, 3.8 and 1.0 million years earlier.

As mentioned in the introduction, a similar calculation was already carried
out in the context of extra-solar planets by Rasio et al.~(1996),
who basically came to the same conclusions; their fig. 2 is reminiscent
of ours. They also employed the orbital decay rate predicted
by Zahn's theory, but their solar evolution model used the old Reimers
mass-loss relation, and they did not make any adjustments to match the
effective temperatures found empirically at the tip of the giant branches
(see Section \ref{evolution}).

Do the remaining uncertainties allow the possibility for
Earth to escape the ``doomsday'' scenario? As far as the mass-loss alone
is concerned, this seems unlikely: according to the study of HB stars
in globular clusters by Schr\"oder \& Cuntz (2005), $\eta$ is remarkably
well constrained and cannot exceed $9 \times 10^{-14} M_\odot\,$y$^{-1}$,
or the total RGB mass-loss would become so large that the tip-RGB star
would miss He-ignition and not reach the horizontal branch at all.
And the full width of the HB towards lower $T_{\rm eff}$ is achieved
already with an $\eta$ of $7 \times 10^{-14} M_\odot\,$y$^{-1}$.
Furthermore, the benefit of larger orbits with a reduced
solar mass is to some extent compensated for by a larger solar giant.

Dynamical drag does not become important until the planet is already
very near the photosphere, i.e., {\it after} tidal drag has already
lowered the orbit. Hence, the most significant uncertainty here comes
from the scaling of the tidal friction coefficient
$\lambda_2$ (of Zahn, 1989). For this reason, we
computed several alternative cases, and from these we find:

(1) With the mass-loss rate unchanged, the value of $\lambda_2$
would have to be significantly smaller for an escape from the
``doomsday'' scenario, i.e.,
less than 0.013, instead of our adopted value of 0.038. But Zahn's scaling
of $\lambda_2$ has been empirically confirmed within a factor of 2, if
not better (see Verbunt \& Phinney, 1995). Very recently, realistic 3D
simulations of the solar convection have also resulted in an effective
viscosity which matches that of Zahn's prescription surprisingly
well (Penev et al. 2007). And Rybicki \&\ Denis (2001), by comparison,
used a value ($K_2 = 0.05$ in the notation of their very similar calculation
of tidal angular momentum loss) which is entirely consistent with Zahn's
scaling of $\lambda_2$ for a convection parameter of $\alpha = 2$.

(2) We then considered solar evolution models with a reasonably larger
mass-loss rate ($\eta = 9 \times 10^{-14} M_\odot\,$y$^{-1}$)
in combination with tidal friction
coefficients of 1/1, 2/3 and 1/2 of the one given by Zahn.
In each of these cases, planet Earth would not be able to escape doomsday
but would face a delayed engulfment by the supergiant
Sun -- 470,000, 230,000 and 80,000 years before the tip-RGB is
reached, respectively.

(3) Finally, we checked the outcome for a reasonably lower mass-loss rate
($\eta = 7 \times 10^{-14} M_\odot\,$y$^{-1}$) in combination with the same
tidal friction
coefficients as above. The engulfment would then happen rather
earlier than with more mass-loss -- 540,000, 380,000 and 270,000 years before
the tip-AGB is reached.

These computations confirm that reducing the solar mass enlarges the planetary 
orbit more than the tip-RGB solar radius, so that the best way to avoid the 
doomsday scenario would be to have as high a mass-loss rate as
possible. However, we believe that the value of $\eta$ in case (2) above 
already is as high as it can be without violating agreement of evolved 
models with observations, and that the smallest value used there for the 
tidal friction coefficient is also at the limits of what is allowed by 
observational constraints. The only possible escape would be if our solar 
giant models were too cool (by over 100\,K in case 2), and therefore 
larger than the real Sun will be. Hence, to avoid engulfment by the tip-RGB 
Sun would require that all three parameters ($\eta$, $\lambda_2$ and 
$T_{\rm eff}$) were at one edge of their uncertainty range, which seems 
improbable. Rather, our computations confirm, with reasonable certainty, 
the classical ``doomsday'' scenario.

\subsection{``Doomsday'' avoidable?}

\begin{figure}
\vspace{60mm}
\includegraphics{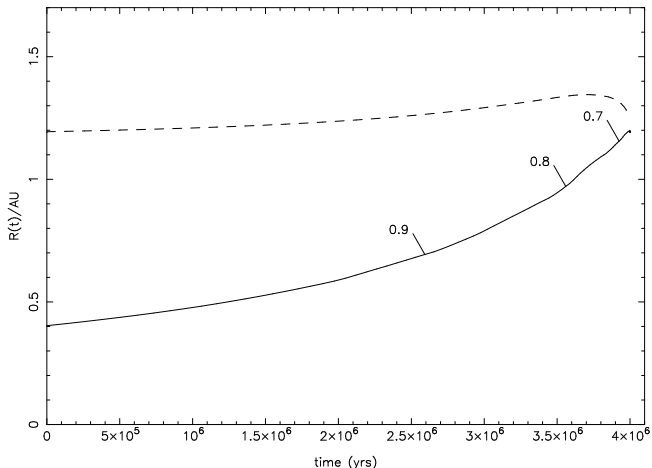}
\caption{As Fig. \ref{drag_earth}, but for a planet with a present orbital
radius of 1.15\,AU.}\label{drag_crit}
\end{figure}

Even though this is an academic question, given the hostile conditions on
the surface of a planet just missing this ``doomsday''scenario, we may ask:
what is the minimum initial orbital radius of a planet in order for it
to ``survive''? Fig. \ref{drag_crit} shows, by the same computation
as carried out for Fig. \ref{drag_earth}, that an initial orbital radius
of 1.15\,AU is sufficient for any planet to pass the tip-RGB of a star
with $M_i = 1.0\,M_{\odot}$. Since, as shown in Section \ref{tipagb}, the
tip-AGB Sun will not reach any similarly large extent again, such a planet
will eventually be orbiting a White Dwarf.

A more general discussion of planetary survival during post-main-sequence
evolution has been given by Villaver \&\ Livio (2007), who suggest that
an initial distance of at least 3\,AU is needed for the survival of a
terrestrial-size planet when one also takes into account the possible
evaporation of the planet by stellar heating. However, they use stellar
models and mass-loss rates that have the maximum radius and mass loss
occurring on the AGB. That has been the expected result for many years,
but is quite different from what we find (Section \ref{tipagb} and Table 1)
with the improved mass-loss formulation of Schr\"{o}der \&\ Cuntz
(2005, 2007). Hence, Villaver \&\ Livio's results may be unduly
pessimistic.

In any case, it is clear that terrestrial planets can survive if
sufficiently far from their parent star. If it were possible to increase
the orbital radius from its initial value, then an increase of only 8\%
of angular momentum should yield the pre-RGB orbital size required by
planet Earth to escape engulfment. Is that conceivable?

An ingenious scheme for doing so which, in the first place,
could increase the time-scale for habitation by intelligent life
for the whole of the Sun's MS life-time, was proposed by Korycansky, Laughlin
\&\ Adams (2001). They pointed out that a suitable encounter of the Earth
every 6000 years or so with a body of large asteroidal mass could be
arranged to move the orbit of the Earth outwards; Kuiper Belt objects
might be the most suitable. The energy requirements could be reduced by
incorporating additional encounters with Jupiter and/or Saturn. Although
still very large by today's standards, the energy requirements remain small
compared to those for interstellar travel.

On the face of it, this scheme seems far-fetched, but Korycansky
et al. (2001) show that it is in principle possible, both technically and
energetically, although currently somewhat beyond our technical capabilities;
however, there is no immediate hurry to implement the scheme, which could
await the development of the relevant technology. It would have the advantage
of improving conditions for the whole biosphere, whereas any
scheme for interplanetary `life rafts' that could move slowly outwards to
maintain habitable conditions would, on cost and energy grounds, necessarily
be confined to a small fraction of the human population -- with all the
political problems that that would produce -- plus perhaps a tiny proportion
of other species.
None the less, the asteroidal fly-by scheme has its own problems,
not least the danger of a benign close approach turning into a
catastrophic accidental collision, and possibly also triggering orbital
instability -- cf. also Debes \&\ Sigurdsson (2002).

\section{Tip-AGB solar evolution}\label{tipagb}
The loss of 1/3 of the solar mass during the rise to the tip of the RGB will
make a significant impact on the further evolution as an AGB star.
There is very little shell mass left, into which the two burning shells
(H, followed by He)
can advance (on a radial mass scale). Hence, the C/O core cannot grow as
much as with a conservative model without mass loss, and the whole core
region will not contract as much, either. Consequently, the AGB luminosity,
determined by the density and temperature in the H-burning shell, will not
reach as high levels as in a conservative AGB model, and neither will the AGB
radius of the late future Sun (see Table \ref{solarmodels}).

\begin{figure}
\vspace{60mm}
\includegraphics{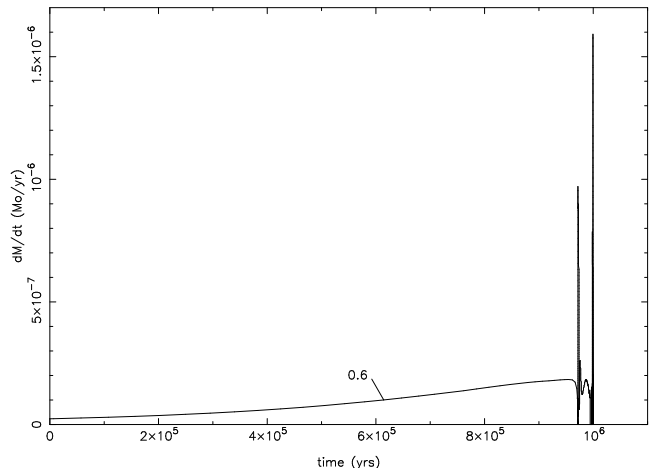}
\caption{Solar mass loss during the final 1 million years on the
AGB will remain mainly of the order of $2 \times 10^{-7} M_{\odot}$\,y$^{-1}$
and not provide sufficient CS shell mass to form a regular PN. Only
the last two TP's (tip-AGB and post-AGB, see text) are resolved.}
\label{mdot}
\end{figure}

According to our evolution model, the regular tip-AGB evolution
will be ended with a luminosity of only $2090\,L_{\odot}$, $T_{\rm eff}
 = 3200$\,K, and $R = 149\,R_{\odot}$. The AGB mass-loss rate, according
to the relation of Schr\"oder \& Cuntz (2005), will reach only
$2.0 \times 10^{-7}\,M_{\odot}$\,y$^{-1}$ (see Fig. \ref{mdot}), since
the luminosity will not be sufficient to drive a dust-driven wind
(see Schr\"oder et al. 1999). Also, even if it did: only a little shell
mass will have been left to lose after the RGB phase, only 0.116$M_{\odot}$.

Hence, for this non-dust-driven AGB solar mass-loss, we have adopted the
same mass-loss description as above (equation~(1)). This mass-loss, in combination
with our solar evolution model, yields the following prediction:
during the final 30,000\,y on the very tip-AGB, which are crucial
for any build-up of sufficient CS (circumstellar) material to form a PN,
the solar giant will lose only $0.006\,M_{\odot}$. A further
$0.0015\,M_{\odot}$ will be lost in just 1300 years right after a
final thermal pulse (TP) on the tip-AGB. That marks the very end of
AGB evolution, and it allows the solar supergiant briefly to reach a
luminosity of $4170\,L_{\odot}$ and $R = 179\,R_{\odot}$,
with a mass-loss rate of $10^{-6}\,M_{\odot}$\,y$^{-1}$, but with $T_{\rm eff}$
already increased to 3467\,K. Again, there will be no involvement of a
dust-driven wind.
Since common PNe and their dusty CS envelopes
reveal a dust-driven mass-loss history of more like $10^{-5}$ to
$10^{-4} M_{\odot}$\,y$^{-1}$ during the final 30,000
years of tip-AGB evolution, we must conclude that the Sun will {\it not}
form such a PN.

Since a circumstellar shell of nearly $0.01\,M_{\odot}$ will, nevertheless,
be produced by the tip-AGB solar giant, a rather peculiar PN may
be created by the emerging hot stellar core -- it might be
similar to IC\,2149. Although most of the peculiar, strongly bi-polar PNe
appear to stem from massive stars, this particular object has only a slim total mass of 0.01 to $0.03\,M_{\odot}$, lacking a massive envelope --
see V{\'a}zquez et al. 2002. Hence, these authors argue that
this PN appears to be the product of a low-mass star with
$M_i$ close to $1\,M_{\odot}$.

A final mass of 0.0036\,$M_{\odot}$ is lost by the post-AGB star, which
on its way to become a hot subdwarf undergoes at least one more TP.
For the resulting solar WD (white dwarf), our evolution model yields a
final mass of $0.5405\,M_{\odot}$.

\section{Conclusions}
We have applied an improved and well-tested mass-loss relation
to RGB and AGB solar evolution models, using a well-tested evolution
code. While the habitable zone in the inner solar system will already
move outwards considerably in the next 5 billion years of solar MS
evolution, marking the end of life on Earth, the most critical and fatal
phase for the inner planetary system is bound to come with the final ascent
of the Sun to the tip of the RGB.

Considering in detail the loss of angular momentum by tidal
interaction and dynamical drag in the lower chromosphere of the solar
giant, we have been able to compare the evolution of the RGB solar radius with
that of the orbit of planet Earth. Our computations reveal that planet Earth
will be engulfed by the tip-RGB Sun, just half a million years before the Sun
will have reached its largest radius of 1.2 AU, and 1.0 (3.8) million years
after Venus (and Mercury) have suffered the same fate. While solar
mass loss alone would allow the orbital radius of planet Earth to grow
sufficiently to avoid this ``doomsday'' scenario, it is mainly tidal
interaction of the giant convective envelope with the closely orbiting planet
which will lead to a fatal decrease of its orbital size.

The loss of about 1/3 of the solar mass already on the RGB has significant
consequences for the solar AGB evolution. The tip-AGB Sun will not qualify
for an intense, dust-driven wind and, hence, will not produce a regular PN.
Instead,
an insubstantial circumstellar shell of just under 1/100\,$M_{\odot}$ will
result, and perhaps a peculiar PN similar to IC\,2149.

\section*{Acknowledgments}
KPS is grateful for travel support received from the Astronomy Centre at
Sussex through a PPARC grant, which enabled the authors to initiate this
research project in the summer of 2006. We further wish to thank
Jean-Paul Zahn for very helpful comments on his treatment of tidal friction
and Adam Scaife of the Met Office's Hadley Centre for suggesting changes
to Sections \ref{intro} and \ref{habitablezone}.

\label{lastpage}

\begin{thebibliography}{}
\bibitem{} Anders E., Grevesse N., 1989, Geochim. Cosmochim. Acta, 53, 197
\bibitem{} Asplund M., Grevesse N., Sauval A. J., 2005, in Barnes III T. G.,
           Bash F. N., eds, Cosmic Abundances as Records of Stellar Evolution
           and Nucleosynthesis, ASP Conf. Ser. 336, Astron. Soc. Pacific,
           San Francisco, p. 25 (2004, astro-ph/0410214)
\bibitem{} Betts R. A., Cox P. M., Collins M., Harris P. P., Huntingford C.,
           Jones C.D., 2004, Theoretical and Applied Climatology, 78, 157
\bibitem{} Cox P. M., Betts R. A., Collins M., Harris P. P., Huntingford C.,
           Jones C.D., 2004, Theoretical and Applied Climatology, 78, 137
\bibitem{} Debes J. H., Sigurdsson S., 2002, ApJ, 572, 556
\bibitem{} Dyck H. M., Benson J. A., van Belle G. T., Ridgway S. T., 1996,
           AJ, 111, 1705
\bibitem{} Eggleton P. P., 1971, MNRAS, 151, 351
\bibitem{} Eggleton P. P., 1972, MNRAS, 156, 361
\bibitem{} Eggleton P. P., 1973, MNRAS, 163, 179
\bibitem{} Franck S., von Bloh W., Bounama C., Steffen M., Sch\"onberner D.,
Schellnhuber H.-J., 2002, in Horneck G., Baumstark-Khan C., eds, Astrobiology:
The Quest for the Conditions of Life, Springer-Verlag, Berlin, p. 47
\bibitem{} Gough D. O., 1981, Solar Phys., 74, 21
\bibitem{} Harrison G., Bingham R., Aplin K., Kellett B., Carslaw K., Haigh J.,
2007, A\&G, 48, 2.7
\bibitem{} Hurley J. R., Pols O. R., Tout C. A., 2000, MNRAS, 315, 543
\bibitem{} Kandel R., Viollier M., 2005, Sp. Sci. Rev., 120, 1
\bibitem{} Kasting J. F., 1988, Icarus, 74, 472
\bibitem{} Kasting J. F., Whitmire D. P., Reynolds R. T., 1993, Icarus, 101, 108
\bibitem{} Korycansky D. G., Laughlin G., Adams F. C., 2001, Ap\&SS, 275, 349
\bibitem{} Laughlin G. P., 2007, Sky \&\ Telescope, June issue, 32
\bibitem{} Lemaire P., Gouttebroze P., Vial J. C., Artzner G. E., 1981,
           A\&A, 103, 160
\bibitem{} Livio M., Soker N., 1984, MNRAS, 208, 763
\bibitem{} Lopez B., Schneider J., Danchi W. C., 2005, ApJ, 627, 974
\bibitem{} Lovelock J., 1979, GAIA -- A New Look at Life on Earth.
           Oxford Univ. Press, Oxford
\bibitem{} Lovelock J., 1988, The Ages of GAIA. W. W. Norton, New York
\bibitem{} Lovelock J., 2006, The Revenge of GAIA. Basic Books, New York
\bibitem{} Maeder A., Meynet G., 1988, A\&AS, 76, 411
\bibitem{} Maltby P., Avrett E. H., Carlsson M., Kjeldseth-Moe O.,
           Kurucz R. L., Loeser R., 1986, ApJ, 306, 284
\bibitem{} Meynet G., Mermilliod J.-C., Maeder A., 1993, ApJ Suppl., 98, 477
\bibitem{} Ostriker E. C., 1999, ApJ, 513, 252
\bibitem{} Penev K., Sasselov D., Robinson F., Demarque P., 2007,
           ApJ, 655, 1166
\bibitem{} Pols O. R., Tout C. A., Eggleton P. P., Han Z., 1995, MNRAS, 274, 964
\bibitem{} Pols O. R., Tout C. A., Schr\"oder K.-P., Eggleton P. P.,
           Manners, J., 1997, MNRAS, 289, 869
\bibitem{} Pols O. R., Schr\"oder K.-P., Hurley J. R., Tout C. A.,
           Eggleton P. P., 1998, MNRAS, 298, 525
\bibitem{} Prialnik D., 2000, An Introduction to the Theory of Stellar
           Structure and Evolution. Cambridge Univ. Press, Cambridge
\bibitem{} Priest E., Lockwood M., Solanki S., Wolfendale A., 2007, A\&G, 48,
           3.7
\bibitem{} Rasio F. A., Tout C. A., Lubow S. H., Livio M., 1996, ApJ, 470, 1187
\bibitem{} Reimers D., 1975, Mem. Soc. Roy. Sci. Li\`{e}ge, 6eme S\'{e}rie,
           8, 369
\bibitem{} Reimers D., 1977, A\&A, 61, 217
\bibitem{} Rybicki K. R., Denis C., 2001, Icarus, 151, 130
\bibitem{} Sackmann I.-J., Boothroyd A. I., Kraemer K. E., 1993, ApJ, 418, 457
\bibitem{} S{\'a}nchez-Salcedo F. J., Brandenburg A., 2001, MNRAS, 322, 67
\bibitem{} Schr\"oder K.-P., 1990, A\&A, 236, 165
\bibitem{} Schr\"oder K.-P., 1998, A\&A, 334, 901
\bibitem{} Schr\"oder K.-P., Cuntz M., 2005, ApJL, 630, L73
\bibitem{} Schr\"oder K.-P., Cuntz M., 2007, A\&A, 465, 593
\bibitem{} Schr\"oder K.-P., Griffin R. E. M., Griffin R. F., 1990,
           A\&A, 234, 299
\bibitem{} Schr\"oder K.-P., Pols O. R., Eggleton P. P., 1997,
           MNRAS, 285, 696
\bibitem{} Schr\"oder K.-P., Winters J. M., Sedlmayr E., 1999,
           A\&A, 349, 898
\bibitem{} Schr\"oder K.-P., Smith R. C., Apps K., 2001,
           A\&G, 42, 6.26
\bibitem{} Svensmark H., 2007, A\&G, 48, 1.18
\bibitem{} Van Belle G. T., Dyck H. M., Benson J. A., Lacasse M. G., 1996,
           AJ, 112, 2147
\bibitem{} Van Belle G. T., Dyck H. M., Thompson R. R., Benson J. A.,
           Kannappan S. J., 1997, AJ, 114, 2150
\bibitem{} VandenBerg D. A., 1991, in James K., ed., The formation and
           evolution of star clusters, ASP Conf. Ser. 13, Astron. Soc.
           Pacific, San Francisco, p. 183
\bibitem{} V{\'a}zquez R., Miranda L. F., Torrelles J. M., Olgu\'{\i}n L.,
Ben\'{\i}tez G., Rodr\'{\i}guez L. F., L\'opez J. A., 2002, ApJ, 576, 860
\bibitem{} Verbunt F., Phinney E. S., 1995, A\&A 296, 709
\bibitem{} Villaver E., Livio M., 2007, ApJ, 661, 1192
\bibitem{} Zahn J.-P., 1977, A\&A, 57, 383
\bibitem{} Zahn J.-P., 1989, A\&A, 220, 112
\end{thebibliography}
\end{document}